\documentclass[reprint,aps,superscriptaddress,nofootinbib]{revtex4-2}

\usepackage{graphicx,amssymb,subcaption}

\usepackage{dsfont,mathrsfs,xcolor,url,verbatim,booktabs}

\usepackage{hyperref}
\hypersetup{colorlinks=true,allcolors=black}
\usepackage[all]{hypcap}
\usepackage{bookmark}

\usepackage{amsmath, braket, array, multirow, float, empheq, tabstackengine, cases}
\setstackgap{L}{1.2\normalbaselineskip}
\setstacktabbedgap{.4em}
\fixTABwidth{T}

\newcommand\ignore[1]{{}}

\DeclareMathOperator{\rank}{rank}
\DeclareMathOperator{\spanop}{span}
\begin{document}

\preprint{Boundary States and the Edge Chain in Semi-Infinite
Two-Dimensional Insulators Iakoub}

\title{Boundary States and the Edge Chain in Semi-Infinite
Two-Dimensional Insulators}
\author{Ilya Iakoub}
\email{ilya.iakoub@umontreal.ca}
\affiliation{Département de physique, Université de Montréal, Montréal, QC, Canada, H3C 3J7}
\author{Nicolas Levasseur}
\email{nicolas.levasseur@umontreal.ca}
\affiliation{Département de physique, Université de Montréal, Montréal, QC, Canada, H3C 3J7}
\author{Richard MacKenzie}
\email{richard.mackenzie@umontreal.ca}
\affiliation{Département de physique, Université de Montréal, Montréal, QC, Canada, H3C 3J7}

\begin{abstract}
    We present a constructive analytical method for finding the edge and corner states of semi-infinite two-dimensional systems. The method consists in first describing the two-dimensional system as a series of one-dimensional unit cells, then constructing a Bloch Hamiltonian and performing an analytic continuation to complex wave vectors. We find complex wave vectors for which the last layer (or the last few layers) of lattice sites decouple from the rest of the system, effectively reducing a two-dimensional problem to a one-dimensional one. We derive a criterion under which the edge and corner states found using our method are adiabatically protected. The boundary states obtained in this way are the modes of the edge chain that cannot be observed from the rest of the lattice: using Shemesh's theorem, equivalent to the observability criterion of control theory, we obtain a necessary and sufficient condition for their existence and identify families of transformations under which the method remains exact. Using our method, the corner states of two-dimensional systems can be understood as edge or defect-bound states of the one-dimensional edge chain.
\end{abstract}

\maketitle

\section{Introduction}\label{sec:intro}

Boundary states are the most direct physical signature of a topological phase, and the bulk-boundary correspondence relates their number to invariants computed from the Bloch Hamiltonian of the infinite, translation-invariant system~\cite{Hatsugai, Ryu_Hatsugai, Ryu_2010, Asboth}. In two dimensions, besides the edge states of Chern insulators~\cite{Haldane, Hatsugai}, higher-order topological insulators host states localized at the corners of a sample~\cite{BBH, Benalcazar2017, Schindler2018}; the two-dimensional extension of the Su-Schrieffer-Heeger (SSH) model~\cite{SSH, LiuWakabayashi2017, SSH_2D, SSH_2D_2, SSH_2D_3} and the breathing kagome lattice~\cite{Ezawa2018, Breathing_Kagome_lattice} are standard examples. Topological invariants predict whether such states exist, but not what they look like, and obtaining their explicit wave functions is often difficult. General analytic frameworks split the eigenvalue problem of the semi-infinite system into bulk equations, solved by Bloch states with complex wave vectors, and boundary equations, which select linear combinations of these states~\cite{Kohn1959, Hatsugai_Riemann, BB_Chern, Dwivedi_Chua, Alase2017, Balabanov2021}. They are powerful, but their application to a given two-dimensional lattice, and in particular to its corners, can be involved.

In some lattices, the boundary states take a much simpler form. Kunst \emph{et al.} showed that destructive interference can make a boundary state vanish exactly on a whole sublattice, so that its wave function follows from the eigenstates of a small subsystem~\cite{Kunst2017}, and used this mechanism to design lattices with exactly solvable edge, hinge and corner states~\cite{Kunst2018, Kunst2019}. In this construction, the existence of exact solutions is guaranteed by the structure of the lattice: each site on which the state must vanish carries a single orbital, so that the destructive-interference conditions can always be satisfied, or a mirror symmetry makes them compatible~\cite{Kunst2019}. These conditions are sufficient, but they do not tell whether exact solutions of this form exist in lattices without this structure, nor what condition such solutions must satisfy in general. This paper addresses both questions.

We first formulate a constructive procedure, which we call the edge chain method. Given a lattice and a choice of the unit cell along its boundary, we impose a zero amplitude on the sites that connect the outermost layer of sites, the ``edge chain,'' to the rest of the lattice, and look for Bloch states with a complex wave vector perpendicular to the boundary that satisfy this constraint. The two-dimensional problem then reduces to that of the edge chain, which can be an extended one-dimensional system in its own right, supplemented by a condition on the complex wave vector. This leads to a simple picture of corner states: the edge chain of a corner is a one-dimensional chain containing a defect, and corner states are its edge states or its soliton-bound states~\cite{SSH, JackiwRebbi, TeoKane2010}. Corner states are thus connected to the well-understood physics of defects in one dimension, in a hierarchy reminiscent of nested Wilson loops~\cite{BBH} and of the ``boundaries of boundaries'' of Ref.~\cite{Kunst2019}.

We then develop the theory behind the method, using the analytically continued Bloch Hamiltonian $h(z)$, where $z$ replaces $e^{ika}$~\cite{Kohn1959, Hatsugai_Riemann, myself}. We split the sites of the unit cell into the set $b$ of sites forced to zero and the set $i$ of the other sites, and write $h(z)=h_1(z)+h_2(z)$, where $h_1$ and $h_2$ are its block-diagonal and block off-diagonal parts. The states produced by the method are then exactly the eigenvectors of $h_i(z)$, the restriction of $h$ to the $i$ sites, that lie in the kernel of the coupling $h_{bi}(z)$ between the two sets.

This formulation turns the search for exact boundary states into a classic problem of linear algebra and control theory: finding the common eigenvectors of two matrices~\cite{Shemesh}. In the language of control theory, the boundary states are the modes of the edge chain that cannot be observed from the rest of the lattice, and they span the unobservable subspace of the pair $(h_{bi}, h_i)$~\cite{Kalman, Hautus, Kailath, Wonham}. Shemesh's theorem then gives a necessary and sufficient condition for the existence of such states, in the form of a rank condition on the observability matrix, and counts them. It also provides a systematic way to characterize the class of systems in which the method applies. We identify families of transformations of the Hamiltonian that leave the number of solutions unchanged: arbitrary modifications of the couplings among the $b$ sites, rescalings of the couplings between the two sets that preserve the kernel of $h_{bi}$, couplings among the $i$ sites mediated by the $b$ sites, which include the effective Hamiltonian of the edge chain, and shifts of the energies of the boundary states. Starting from any system in which the method works trivially, such as a bipartite lattice with chiral symmetry, these transformations generate systems without chiral symmetry whose boundary states are still given exactly by the edge chain, at energies that need not vanish.

The same formulation also gives a criterion for protection. In a previous work~\cite{myself}, we showed that, in chiral chains, Bloch states vanishing on one sublattice arise at exceptional points of $h(z)$~\cite{Nimrod_Moiseyev, Kato, Higher_order_E.P., non_Hermitian}, so that they can only disappear from the semi-infinite chain through a closing of the bulk gap. We show that this mechanism carries over to the edge chain method precisely when $h_1$ acts identically on the two states that coalesce at the exceptional point of $h_2$, and that the states found by the method are not protected otherwise. This explains, within a single framework, why the boundary states of the two-dimensional SSH model are robust while those of the trimer SSH model~\cite{SSH3} and of zigzag graphene with a staggered sublattice potential~\cite{Fujita1996, Semenoff1984} can disappear without any closing of the gap.

The method relies on nearest-neighbor hoppings across the boundary, which is what allows a single Bloch state to satisfy the boundary condition. Longer-range hoppings impose additional boundary conditions, and the boundary states must then be built from linear combinations of Bloch states~\cite{Alase2017, myself}; the framework developed here makes this limitation explicit.

The paper is organized as follows. In \autoref{sec:Edge Chain Ansatz}, we introduce the method on the two-dimensional SSH model, and obtain both its corner and its edge states. In \autoref{sec:chiral quasi symmetry}, we restate the method in terms of the analytically continued Bloch Hamiltonian, and derive the condition under which the boundary states it produces can only disappear through a gap closing. In \autoref{sec:range of validity}, we use Shemesh's theorem to characterize the systems in which the method applies. In \autoref{sec:Beethoven}, we interpret corner states as boundary states of the edge chain, and \autoref{sec:Conclusion} contains our concluding remarks. The appendices apply the method to the trimer SSH model (\hyperref[app:SSH3]{App.~\ref{app:SSH3}}), which also shows that it applies in one dimension, to zigzag graphene with a staggered potential (\hyperref[app:Haldane]{App.~\ref{app:Haldane}}), and to the breathing kagome lattice (\hyperref[app:Kagome]{App.~\ref{app:Kagome}}), where we recover, within our framework, the exact corner states obtained in Refs.~\cite{Kunst2018, Kunst2019}.

\section{Edge Chain Method}\label{sec:Edge Chain Ansatz}

In this paper, we are proposing a method for determining the edge and corner states of semi-infinite two-dimensional systems. We will start by working through an example to illustrate our method. In the following sections, we find the conditions under which this method will yield an edge or corner state.

\begin{figure}[t]
    \centering
    \includegraphics[width=1.0\linewidth]{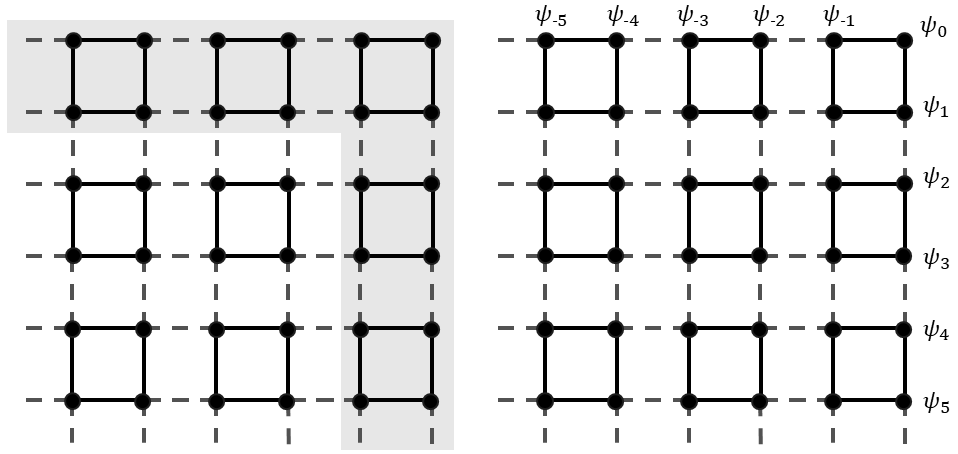}
    \caption{2D extension of the SSH model. The full lines are $t_1$ hoppings, the dashed lines are $t_2$ hoppings. In the shaded area, the unit cell we have used is illustrated. On the right, we show the labeling of the edge chain sites.}
    \label{fig:SSH 2D}
\end{figure}

Consider a semi-infinite two-dimensional extension of the SSH model~\cite{SSH_2D, SSH_2D_2, SSH_2D_3} shown in \autoref{fig:SSH 2D}. We choose a semi-infinite unit cell along the boundary of the sample. Let us label the hopping parameters with $t_1$ and $t_2$ (both taken to be real and positive without loss of generality), and the amplitude of the sites along the boundary using $\psi_i$, with $i>0$ corresponding on the right boundary, $i<0$ corresponding on the sites on the top boundary and $\psi_0$ corresponding to the amplitude of the site on the corner (see \autoref{fig:SSH 2D}). We call the chain of sites along the boundary of the system the ``edge chain.” Consider only the edge of the sample: it describes an infinite one dimensional SSH model with a (perfectly localized) soliton in its center, corresponding to the corner. We can solve the one-dimensional chain~\cite{SSH_with_Soliton, Nicolas} to obtain a set of ``bulk” (oscillating) states and one soliton-bound (exponential) state.

\begin{figure}[t]
    \centering
    \includegraphics[width=0.6\linewidth]{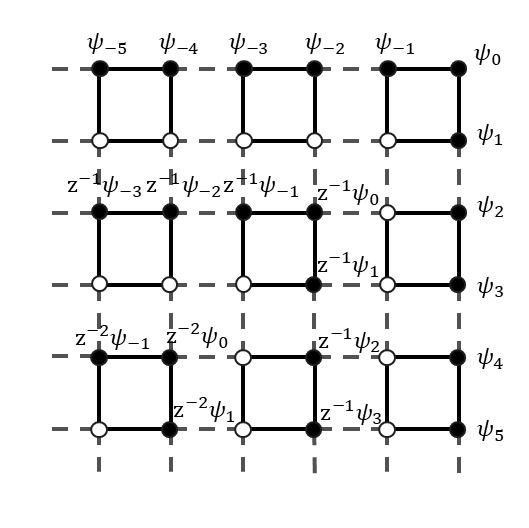}
    \caption{To enforce the edge chain \textit{ansatz}, we set every second ``chain” to zero. The resulting wave function is shown above. The white circles represent sites where the wave function amplitude is zero.}
    \label{fig:SSH 2D avec zeros}
\end{figure}

The crux of the method we are proposing is as follows\,: use these states as an \textit{ansatz} for solving the full system. Clearly, if the boundary is not coupled to the rest of the system, the solutions that we found will be exact, so we may ask ourselves: how can this happen? To effectively decouple the boundary from the rest of the system, we can force every site neighboring the edge chain to have amplitude zero (this works because there are no hoppings beyond those to nearest neighbors). To do so, we can copy the amplitudes $\psi_i$ to the next unit cells in such a way that the hoppings cancel just in the right spots. This is illustrated in \autoref{fig:SSH 2D avec zeros}. Here, $z$ essentially corresponds to $e^{ika}$, with $a$ being the lattice constant, but we allow $k$ to have an imaginary component. Notice that such a state is a pure Bloch state for the unit cell shown in \autoref{fig:SSH 2D}. By applying the Hamiltonian on the sites we want fixed to zero, we obtain the conditions
\begin{align}
    t_1 \psi_{i-2}+z^{-1}t_2\psi_i=0 \mspace{10mu}\text{for $i\leq0$} \\
    t_1 \psi_{i+2}+z^{-1}t_2\psi_i=0 \mspace{10mu}\text{for $i\geq0$}\\
    t_1(\psi_1 + \psi_{-1})=0,
\end{align}
which implies $z=-\frac{t_2 \psi_i}{t_1 \psi_{i-2}}$ if $i\leq 0$; and $z=-\frac{t_2 \psi_i}{t_1 \psi_{i+2}}$ if $i\geq 0$. Finally, since $z$ cannot depend on the lattice site, we must have $\frac{\psi_i}{\psi_{i-2}}=cst.$ for $i\leq 0$ and $\frac{\psi_i}{\psi_{i+2}}=cst.$ for $i\geq 0$; therefore $\psi_{i-2}=\alpha \psi_i$ in $i\leq 0$ and $\psi_{i+2}=\alpha \psi_i$ in $i\geq 0$. The only states satisfying those conditions are
\begin{equation}
 \psi_i =
 \begin{cases}
     (-\frac{t_1}{t_2})^{|i/2|} \text{ for $i$ even}\\
     0 \text{ for $i$ odd}
 \end{cases},
\end{equation}
if $t_1<t_2$, and
\begin{equation}
    \psi_i=
    \begin{cases}
        0\text{ for $i$ even}\\
       \text{sgn}(i)\, (-\frac{t_2}{t_1})^{(|i|-1)/2} \text{ for $i$ odd}
    \end{cases},
\end{equation}
if $t_1>t_2$. As a result, $z=\frac{t_2^2}{t_1^2}$ in the first case, and $z=1$ in the second (which does not decay into the bulk, it is therefore not a boundary state). The first state is the corner state of the 2D SSH model.

\begin{figure}[t]
    \centering
    \includegraphics[width=1.0\linewidth]{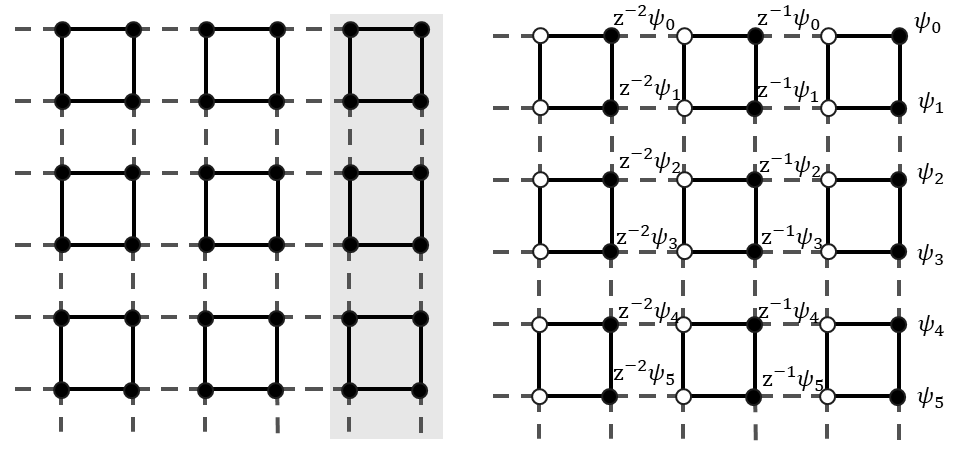}
    \caption{The 2D SSH model with another unit cell identified by a shaded area in the image on the left. On the right, we show the corresponding \textit{ansatz}. The white circles represent sites where the wave function amplitude is zero.}
    \label{fig:2DSSH avec autre cellule unitaire}
\end{figure}

The two-dimensional SSH model also has states localized at its edges rather than at its corners, which our \textit{ansatz} failed to take into account. We can, however, find them using the same method simply by changing the unit cell we are using. Consider the unit cell shown in \autoref{fig:2DSSH avec autre cellule unitaire}. We now have the condition
\begin{equation}
    t_1\psi_i+z^{-1}t_2\psi_i=0,
\end{equation}
which does not imply anything on $\psi_i$. In fact, if $t_1<t_2$, any solution of the chain on the edge (this time without a soliton) will be a solution of the full system, as long as $z=-\frac{t_2}{t_1}$ (if $t_1>t_2$, there will be exponential growth towards the bulk). The same process can be repeated for the upper edge. By choosing one unit cell rather than the other, we restrict the directions in which the wave function is allowed to grow. The first unit cell we used assumed exponential growth towards the corner, prohibiting us from finding the edge states, whereas the second  unit cell assumed exponential growth towards one edge, prohibiting us from finding any other boundary localized state.

Although this model in particular can be solved exactly in a pretty straightforward manner,\footnote{$H$ can be written as $H= I \otimes H_{1D\mspace{2mu}SSH}+H_{1D\mspace{2mu}SSH}\otimes I  $ ($H_{1D\mspace{2mu}SSH}$ is the one dimensional semi-infinite SSH Hamiltonian), which allows to construct 2D SSH eigenstates from the 1D ones.} the method we proposed here allows to analytically find edge and corner states in more complicated models (see \hyperref[app:SSH3]{appendices}) and even in cases where the edge is not a simple straight line. We can summarize the method like this:
\begin{enumerate}
    \item Find the directions in which the wave function is allowed to grow exponentially. Choose a unit cell along a boundary which may host edge states. \label{item:unit cell}
    \item Consider the chain on the surface of the system, i.e.the last layer of sites. Assume that sites that connect the edge chain to the bulk have zero amplitude. \label{item:zero amp}
    \item Assume the solution is a Bloch state with complex wave number $z=e^{ika}\in\mathbb{C}$ and with the unit cell found in step~\ref{item:unit cell}. \label{item:bloch}
    \item Applying the Hamiltonian operator to the sites fixed to zero in step~\ref{item:zero amp}, find the restrictions imposed on the wave function of the edge chain. \label{item:restrictions}
    \item Solve the edge chain. Solutions which also satisfy the conditions established in step~\ref{item:restrictions} will correspond to edge and corner states. \label{item:solve}
\end{enumerate}
We will hereafter refer to this set of steps as “the method.” The method is constructive: by applying the steps above, one constructs an edge state. Nothing, so far, guarantees that the method works, but when it does the output is always a valid edge state.

Note that, in general, what we call ``the edge chain” may in fact be a ribbon (finite width, infinite length). In the following, we will use the words ``ribbon” and ``chain” interchangeably.

The next few sections will be dedicated to justifying why we can expect our \textit{ansatz} to work and understanding its mechanisms.

\section{Mathematical statement of the method}\label{sec:chiral quasi symmetry}

Let us investigate a semi-infinite lattice with nearest-neighbor hoppings through the Hamiltonian of the corresponding infinite lattice. As in Sec.~\ref{sec:Edge Chain Ansatz}, we choose a unit cell along the edge of the semi-infinite system, so that the semi-infinite system is obtained from the infinite lattice by removing every unit cell beyond the edge.

We then split the sites of the unit cell into two sets. The set $b$ contains the sites which, in the first removed unit cell, are connected by a hopping to the semi-infinite system; the set $i$ contains all the other sites of the unit cell. Because the unit cell is repeated periodically, each set also defines a sublattice of the whole infinite lattice. In this basis, the Hamiltonian reads
\begin{equation}\label{eq:Separation of Hamiltonian}
    H=H_1+H_2,
\end{equation}
with
\begin{equation}
    H_1=
    \begin{pmatrix}
        H_{b} & 0 \\
        0 & H_{i}
    \end{pmatrix}
    ;\mspace{20mu}H_2=
    \begin{pmatrix}
        0&H_{bi}\\
        H_{ib}& 0
    \end{pmatrix},
\end{equation}
where $H_b$ ($N_b \times N_b$) and $H_i$ ($N_i \times N_i$) contain the hoppings within each sublattice, and $H_{bi}=H_{ib}^\dagger$ the hoppings between them. Since $b$ and $i$ are defined unit cell by unit cell, $H_b$, $H_i$ and $H_{bi}$ are themselves periodic.

\begin{figure}[t]
    \centering
    \includegraphics[width=1.0\linewidth]{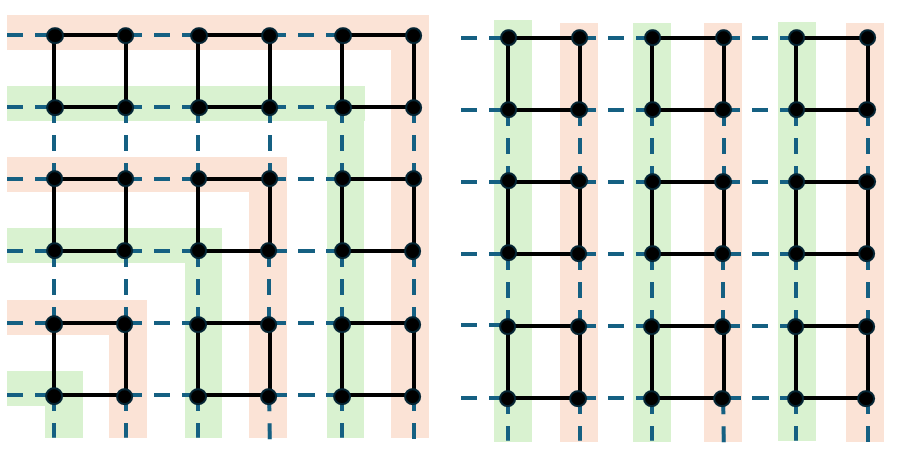}
    \caption{(Color online). In orange: sites with amplitude $\psi_i$; in green sites with amplitude $\psi_b$. When a Bloch state satisfies the boundary condition, $\psi_b=0$. On the left: the $\psi_i$ and $\psi_b$ for the unit cell in \autoref{fig:SSH 2D}; on the right: for the unit cell in \autoref{fig:2DSSH avec autre cellule unitaire}.}\label{fig:labelling}
\end{figure}

The point of this choice is the following. Take a Bloch state of the infinite lattice that vanishes on $b$, that is, of the form $\begin{pmatrix} 0 & \psi_i \end{pmatrix}^T$. Since it is a Bloch state, it vanishes on the $b$ sites of every unit cell, in particular on the removed sites adjacent to the edge. With nearest-neighbor hoppings, the removed part of the lattice then exerts no influence on the semi-infinite system, and the restriction of this state to the semi-infinite system is an eigenstate of it. Steps~\ref{item:unit cell} and~\ref{item:zero amp} of the method are therefore equivalent to the single condition $\psi_b=0$. This partition is illustrated for the two-dimensional SSH model in \autoref{fig:labelling}: the green sites form $b$ and the orange sites form $i$.

\subsection{Bloch Hamiltonian and Analytic Continuation}
This system has translational invariance and thus we can describe it with a Bloch Hamiltonian
\begin{equation}
    h(k) = h_{1}(k)+h_{2}(k),
\end{equation}
with 
\begin{equation}
    h_{1}(k)=
    \begin{pmatrix}
        h_{b}(k) & 0 \\
        0 & h_{i}(k)
    \end{pmatrix}
\end{equation}
and 
\begin{equation}
    h_{2}(k)=
    \begin{pmatrix}
        0 & h_{bi}(k)\\
        h_{ib}(k) & 0
    \end{pmatrix},
\end{equation}
in the same basis as (\ref{eq:Separation of Hamiltonian}). Step~\ref{item:bloch} requires the Bloch Hamiltonian to be analytically continued. This analytic continuation is done in the following way~\cite{myself}: let $M$ be the farthest hopping in terms of unit cells and $|p;\mu\rangle$ be the $\mu$-th site of the $p$-th unit cell. The components of the analytically continued Bloch Hamiltonian are
\begin{equation}
    (h(z))_{\mu,\mu'} = \sum_{m=-M}^M z^{-m} t^{m}_{\mu,\mu'},
\end{equation}
where $t^m_{\mu,\mu'}$ are hopping amplitudes to the $m$-th neighboring unit cell, from a sublattice $\mu$ to a sublattice $\mu'$ (see~\cite{myself} for more details). Note that $h(e^{ika})$ yields the Bloch Hamiltonian. Several important points are worth noting about $h(z)$. Firstly, $h(z)$ is not, in general, Hermitian; it is only Hermitian for $|z| = 1$. Secondly, $h(z)=(h(1/z^*))^\dagger$. This property is called para-Hermiticity~\cite{para_Hermitian_1, para_Hermitian_2}. It implies that the right and left eigenvectors of $h(z)$ satisfy 
\begin{equation}\label{eq:para-Hermitian scalar product}
    \langle m,1/z^*|n, z\rangle = N_m\delta_{m,n},
\end{equation}
where $m$ and $n$ are band indices and $N_m$ is the normalization.

With this continuation, the Bloch Hamiltonian becomes
\begin{align}
    &h(k) \to h(z) = h_1(z) + h_2(z)\\
    &=
    \begin{pmatrix}
        h_{b}(z) & 0 \\
        0 & h_{i}(z)
    \end{pmatrix}
    +
    \begin{pmatrix}
        0 & h_{bi}(z)\\
        \big(h_{bi}(1/z^*)\big)^\dagger & 0
    \end{pmatrix}.
\end{align}

The eigenstates of $h(z)$ are described by
\begin{equation}
    \alpha(z)=
    \begin{pmatrix}
        \alpha_{b}(z)\\
        \alpha_{i}(z)
    \end{pmatrix}
\end{equation}
and states with $\alpha_{b}(z)=0$ are the ones we are looking for in the method.

In terms of these states, the method can be stated as
\begin{align}
    h(z)
    \begin{pmatrix}
        0\\
        \alpha_i(z)
    \end{pmatrix} = 
    \begin{pmatrix}
        h_{bi}(z)\alpha_i(z)\\
        h_i(z)\alpha_i(z)
    \end{pmatrix} = E \begin{pmatrix}
        0\\
        \alpha_i(z)
    \end{pmatrix},
\end{align}
that is we are looking for a state $\begin{pmatrix}
    0 & \alpha_i
\end{pmatrix}^T$ such that $\alpha_{i}(z) \in \ker h_{bi}(z)$ and is an eigenstate of $h_{i}(z)$. If this happens at $|z|\neq 1$, this state is an edge state. Note that, if that state is going to be legitimate, there must be an additional requirement that either $|z|\geq 1$ or $|z|\leq 1$, depending on the boundary. Otherwise, the state may not be normalizable.

$h_{2}(z)$ possesses chiral symmetry, with the chiral symmetry operator
\begin{equation}
    \Gamma =\begin{pmatrix}
     I  & 0\\
    0 & - I 
\end{pmatrix}.
\end{equation}
Note that this is not a physical symmetry of the system, only of $h_2(z)$. Let us consider an eigenstate of $h_{2}(z)$,
\begin{equation}
    \beta^{\pm}(z) = \begin{pmatrix}
        \beta_{b}(z)\\
        \pm\beta_{i}(z)
    \end{pmatrix}.
\end{equation}
The $\beta_{b}(z)=0$ states can arise at exceptional points of $h_{2}(z)$~\cite{myself}. At those points, the non-Hermitian, analytically continued, Bloch Hamiltonian $h_2(z)$ has two coalescing bands. In one dimensional chains with nearest-neighbor hopping, such points in a chiral symmetric Hamiltonian correspond to adiabatically protected edge states~\cite{myself}. In the next subsection, we will explore how this observation applies to two-dimensional systems.

\subsection{Conditions for Protection}\label{sec:band crossing}

Consider $|\beta_{\pm}\rangle$, a positive (negative) energy eigenstate of $h_2(z)$. If $|\beta_{+}\rangle$ coalesces with $|\beta_{-}\rangle$, $h_2(z)$'s positive and negative energy bands touch at the exceptional point. Now imagine that $h_2(z)$ is continuously deformed, $z_0$ becomes a mobile point. If the trajectory of $z_0$ touches the unit circle, where $z_0 = e^{ik_0 a}$, the bands touch at a real value of $k$ and there must be a gap closing in the Brillouin zone. For a given boundary, the edge states can only exist for $|z|>1$ or $|z|<1$, therefore if an edge state disappears, it must cross the unit circle and close the gap of $h_2$~\cite{myself}. The exceptional states of $h_2$ are therefore protected, at least as long as the exceptional points of $h_2$ do not disappear. This is guaranteed as long as $h_2$ remains chiral symmetric with respect to the same operator $\Gamma$, i.e., as long as it remains block off-diagonal in the original partition: a generic transformation mixes the $b$ and $i$ blocks and breaks this property.

The full Hamiltonian, however, is not limited to $h_2$, it is given by $h(z)=h_1(z)+h_2(z)$ and $h_2$ is simply its block off-diagonal part. We would like to see for which $h_1$ does the gap closing argument above extend. Namely we are interested in the shared eigenstates of $h_1$ and $h_2$ where the state is also at an exceptional point of $h_2$. Let us start by considering a generic two band model. Its analytically continued Bloch Hamiltonian is always of the form
\begin{equation}
    h(z)=
    \begin{pmatrix}
        a(z) & b(z)\\
        b^*(1/z^*) & d(z)
    \end{pmatrix},
\end{equation}
and at the exceptional point of $h_2$, $z_0$, either $b^*(1/z_0*)=0\neq b(z_0)$ or $b(z_0)=0\neq b^*(1/z_0^*)$. Without loss of generality, consider the latter case. When $b(z_0)=0$, the eigenvalues of $h(z)$ are $E \in \{a(z),d(z)\}$, and the bands only meet at a given point for $a(z)=d(z)$. Particularly, on the unit circle, $b^{*}(z)=b^{*}(1/z^*)$ and both vanish identically, thus, $h$ becomes diagonal with energies $a$ and $d$. The gap at that point is $|a-d|$.

To generalize this argument, we note that the key insight from the treatment above is that the gap of $h(z)$ closes at $|z_0|=1$ when $h_1(z)$ acts identically on the two coalescing states. The eigenstates of $h_2(z)$ are of the form
\begin{equation}
    \beta^{\pm}=
    \begin{pmatrix}
        \beta_b\\
        \pm \beta_i
    \end{pmatrix}.
\end{equation}
Therefore $h_1 \beta^{\pm} = \lambda \beta^{\pm}$ (with the same $\lambda$) if and only if $h_1 (\beta^{+} \pm \beta^{-})=\lambda (\beta^{+} \pm \beta^{-})$, from which it follows that
\begin{align}\label{eq:exceptional point condition}
    h_i\beta_i&=\lambda \beta_i\\
    h_b\beta_b&=\lambda \beta_b\nonumber.
\end{align}
For example, in the 2D SSH model partitioned according to the right of \autoref{fig:labelling}, $h_b=h_i$ and $\beta_i\propto\beta_b$, therefore the bands of $h(z)$ meet at the exceptional point and we can expect the edge states to be protected as long as the structure of $h(z)$ is preserved. In particular, an evolution that preserves these edge states must keep the off-block-diagonal form of $h_2$ and must keep the way $h_1$ acts on the coalescing states.

We state the following sufficient condition for adiabatically protected edge states in two-dimensional lattices with nearest-neighbor hoppings. Consider a continuous family $h(z,t)=h_1(z,t)+h_2(z,t)$, $t\in[0,1]$, such that $h(z,0)$ has an edge state, found by the method, arising at an exceptional point of $h_2(z,0)$. Suppose that, for all $t$, (i) this state remains an exceptional state of $h_2(z,t)$ and (ii) \eqref{eq:exceptional point condition} remains satisfied.
Then the edge state can only disappear by crossing the unit circle, where two bands of $h$ meet. As long as the gap of $h(e^{ika},t)$ remains open, the edge state therefore persists.

Generically, an exceptional state of $h_2$ that is also an eigenstate of $h_1$ does not lead to two bands of $h$ meeting at the exceptional point and such states are not protected.

\section{Range of Validity of the Method}\label{sec:range of validity}

We would like to understand the range of validity of the method. To do so, we consider a Hamiltonian where the method works and find a set of transformations that preserves the validity of the method. Equivalently, we find a set of transformations that leave the number of eigenstates shared between $H_2$ and $H_1$ (with $\psi_b=0$) unchanged.\\

As usual, we write eigenstates of $H$ as
\begin{equation}
    \psi =
    \begin{pmatrix}
        \psi_b\\
        \psi_i
    \end{pmatrix}
\end{equation}
and Bloch states satisfying the boundary condition in a system with nearest-neighbor hoppings will have $\psi_b=0$. Therefore, the states with which we are concerned must also satisfy
\begin{align}
    H_{bi}\psi_i&=0 \label{eq:conditions1} \\
    H_i\psi_i&=E_i\psi_i, \label{eq:conditions2}
\end{align}
which means they are in the kernel of $H_{bi}$ and are eigenstates of $H_i$. We want to know how to count the number of simultaneous solutions of (\ref{eq:conditions1}) and (\ref{eq:conditions2}) in order to understand the transformations that leave this number unchanged. It so happens that this exact problem is known in the context of control theory~\cite{Wonham, Hautus, Kalman, Kailath}.\\

\textbf{Theorem (Shemesh, 1984)}~\cite{Shemesh}: There is an eigenvector $\psi_i$ of $H_i\in \mathbb{C}^{N_i \times N_i}$ which satisfies $H_{bi}\psi_i=0$, where $H_{bi}\in\mathbb{C}^{N_b \times N_i}$, if and only if $\mathcal{M}=\cap_{l=0}^{\infty}\ker(H_{bi}H_i^l)\neq \{0\}$.

Furthermore, if the restriction of $H_i$ to $\mathcal{M}$ is diagonalizable, the number of linearly independent solutions to the system of equations (\ref{eq:conditions1}) and (\ref{eq:conditions2}) is equal to $\dim \mathcal{M}$~\cite{Shemesh}. Following Shemesh, we can define the matrix
\begin{equation}\label{eq:L}
    L\equiv
    \begin{pmatrix}
        H_{bi}\\
        H_{bi}H_i\\
        H_{bi}H_i^2\\
        \vdots\\
        H_{bi}H_i^{q-1}
    \end{pmatrix}
\end{equation}
where $q$ is the degree of the minimum polynomial of $H_i$ (note that taking $q$ to infinity does not affect the result). This allows us to write $\dim \mathcal{M}=N_i- \rank L$. In practice, we apply the theorem to the Bloch Hamiltonian at fixed $z$ (and, in two dimensions, at fixed momentum along the edge), where $h_i(z)$ and $h_{bi}(z)$ are finite matrices. The computation of $\rank L$ gives important information about the number of edge states.\\

The task of finding the transformations of the Hamiltonian that leave the number of states satisfying (\ref{eq:conditions1}) and (\ref{eq:conditions2}) unchanged is equivalent to finding transformations of the Hamiltonian that preserve the rank of $L$. In control theory, the matrix $L$ is known as the observability matrix of the pair $(H_{bi},H_i)$, and subspace $\mathcal{M}$ is known as the unobservable subspace.

Now, let us find the exact transformations that preserve the number of eigenvectors of $H_i$ in the kernel of $H_{bi}$. Part of this question has an established answer coming from control theory. Let $\mathcal{M}$ be the ``unobservable subspace" of $(C,A)$ (in our context $A=H_{i}$ and $C=H_{bi}$), the dimension of $\mathcal{M}$ is preserved by transformations of the form $A\to A+KC$~\cite[Lemma 2.1]{Wonham} and $C \to VC$. Physically this allows for the first three of these five families of transformations:
\begin{enumerate}
    \item $H_b \to H_b'$. Only the internal couplings of the $b$-subspace (see green subspace in \autoref{fig:labelling}) are changed. One can see this transformation is allowed from the fact that \eqref{eq:conditions1} and \eqref{eq:conditions2} do not involve $H_b$ at all. Notably, one could add disorder to the $b$-subspace and the method would still work.
    
    \item $H_{bi} \to VH_{bi}$ with $V$ injective on the image of $H_{bi}$. These transformations follow from the fact that, for such an injective $V$, $VH_{bi}$ and $H_{bi}$ have the same kernel, therefore \eqref{eq:conditions1} remains unaffected. These transformations change the couplings between the $i$ and $b$ subspaces in such a way that the wavefunction still cancels on the $\psi_b$ subspace (green subspace in \autoref{fig:labelling}).
    
    \item $H_i \to H_i + H_{bi}^\dagger Z H_{bi} $ for $Z$ hermitian. For these transformations, \eqref{eq:conditions2} is satisfied because $H_{bi}\psi_i=0$. $Z$ acts effectively as a propagator: it couples the $i$ subspace to itself by sending $i \to b \to i$. Notably, if $Z = (E-H_{b})^{-1}$, $H_i'$ becomes the effective Hamiltonian for the $i$ subspace.
    
    \item $H_i \to UH_i U^\dagger$ and $H_{bi} \to H_{bi}U^\dagger$ with $U$ unitary. This corresponds to a change of basis of $i$ subspace, therefore it cannot affect the number of simultaneous solution of \eqref{eq:conditions1} and \eqref{eq:conditions2}. A change of basis of the $b$ subspace is also, obviously allowed.

    \item $H_i \to H_i + D$ where $D$ is Hermitian and $[D,H_i]=0$. In that case, $D$ and $H_i$ share a set of eigenvectors, and the transformation only shifts their energies. If $H_i$ is degenerate, there is the additional condition that $D \mathcal{M}=\mathcal{M}$.

\end{enumerate}

Note that rules 1 to 4 preserve the energies of the edge states. If the method works for a given system, it also works for every system to which the given system is related by a combination of these five transformations. One can take a Hamiltonian where the method works trivially (for example one where $H_1=0$), vary it according to the rules we established above and get another Hamiltonian where the method works.

For instance, one can start from a bipartite nearest-neighbor model, such as the SSH chain or graphene with a zigzag edge, with $b$ and $i$ the two sublattices. Then $H_b=H_i=0$ and the method is exact by chiral symmetry. Transformations 1 to 3 add couplings within the $b$ sublattice, rescale the couplings between sublattices, and add couplings within the $i$ sublattice that go through the $b$ sites, while transformation 4 changes the basis within each sublattice. Transformation 5 then shifts the energies of the edge states, as a uniform on-site energy on the $i$ sublattice does in the SSH chain with a staggered potential. The resulting Hamiltonians have no chiral symmetry, yet their edge states remain exact eigenstates, now at energies that need not vanish.

Note that these transformations preserve the validity of the method but do not necessarily keep the gap open. This is not a statement about the robustness of the edge states, but about the systems in which the method yields an edge state.

\section{Corner states as boundary and defect-bound states of the edge chain}\label{sec:Beethoven}

Let us consider a system in which the edge chain can be decoupled from the rest, using our procedure, for some value $z_0$ of the parameter $z$. Throughout this section, we will always assume that $|z_0|< 1$ or $|z_0|>1$ in such a way that the edge chain wavefunction always decays towards the bulk.

At $z_0$, the edge chain is entirely described by the Hamiltonian $h_i(z_0)$. It is often the case that the edge chain, described by $h_i(z_0)$, can itself host edge states. In particular, $h_i(z_0)$ can describe a one-dimensional topological insulator. Take for example the two-dimensional SSH model considered in \autoref{sec:Edge Chain Ansatz}. For the unit cell in \autoref{fig:2DSSH avec autre cellule unitaire}, $h_i(z_0)$ is itself an ordinary SSH chain, which is known to have edge states. In the two-dimensional system, those ``edge states of edge states'' are zero-dimensional objects: corner states. It may also happen, for some unit cells, that the edge chain has a kink (see \autoref{fig:SSH 2D avec zeros}). This is the case when the boundary of the two-dimensional system is not a straight line, notably when it contains a corner. In that case, $h_i(z_0)$ describes an infinite chain with a defect that is localized at a point corresponding to the corner of the two-dimensional system. If $h_i(z_0)$ is a topological chain, a localized state will be bound to the defect~\cite{Nicolas, SSH, JackiwRebbi, TeoKane2010}. From this point of view, corner states of a two-dimensional system can be understood as boundary or defect-bound states of the edge chain. This explains, for example, the corner states in the already discussed two-dimensional SSH model, and in the breathing kagome lattice, shown in \hyperref[app:Kagome]{App.~\ref{app:Kagome}}.  

\section{Conclusion}\label{sec:Conclusion}

We have introduced the edge chain method, which reduces the computation of boundary states of semi-infinite two-dimensional lattices to that of a one-dimensional edge chain. Formulated with the analytically continued Bloch Hamiltonian, the boundary states it produces are the modes of the edge chain that cannot be observed from the rest of the lattice. Shemesh's theorem then gives a necessary and sufficient condition for their existence, and identifies transformations, including ones that break chiral symmetry, under which the method remains exact. These states can only disappear through a gap closing when $h_1$ acts identically on the states that coalesce at the exceptional point of $h_2$. Corner states appear as boundary or defect-bound states of the edge chain, a hierarchy similar in spirit to the ``boundaries of boundaries'' of Ref.~\cite{Kunst2019} and to the nested Wilson loops of Ref.~\cite{BBH}, which suggests a possible connection between these constructions.

Extending the method to longer-range hoppings across the boundary, which require combinations of Bloch states, and to three-dimensional systems are natural directions for future work.

\begin{acknowledgments}
We thank Kylian Lionnet, Hichem Eleuch, Michael Hilke and Andrew Mckenna for useful conversations. This work was supported in part by the Natural Science and Engineering Research Council of Canada and by the Fonds de Recherche Nature et Technologies du Qu{\'e}bec via the INTRIQ strategic cluster grant.
\end{acknowledgments}

\appendix

\section{SSH3 Model}\label{app:SSH3}

\begin{figure}[t]
    \centering
    \includegraphics[width=1.0\linewidth]{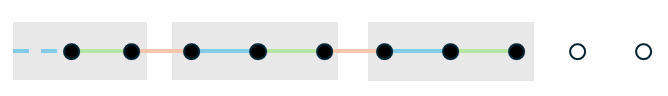}
    \caption{(Color online.) The SSH3 model. In blue: $t_1$ hopping, in green: $t_2$ hopping and in orange: $t_3$ hopping. The unit cell is highlighted in gray. The white circles represent zero amplitude sites.}
    \label{fig:SSH3}
\end{figure}

We want to illustrate that the method also applies in one dimension. To do so, we will use the SSH3 model, shown in \autoref{fig:SSH3}, as an example.

The Bloch Hamiltonian of this model can be written as
\begin{equation}
    h(z)=
    \begin{pmatrix}
        0 & t_1 & t_3 z^{-1}\\
        t_1 & 0 & t_2\\
        t_3 z & t_2 & 0
    \end{pmatrix}.
\end{equation}
Suppose that the boundary arises between the first and second sites of a unit cell. In that case, we have the following splitting of the Hamiltonian:
\begin{equation}
    h_{1}(z)=
    \begin{pmatrix}
        0 & 0 & 0\\
        0 & 0 & t_2\\
        0 & t_2 & 0
    \end{pmatrix};
    \mspace{20mu}
    h_{2}(z)=
    \begin{pmatrix}
        0 & t_1 & t_3 z^{-1}\\
        t_1 & 0 & 0\\
        t_3 z & 0 & 0
    \end{pmatrix}.
\end{equation}
We are interested in 
\begin{equation}
    \ker h_{bi}(z)=\ker \begin{pmatrix}
    t_1 & t_3 z^{-1}
\end{pmatrix}= \spanop \{
\begin{pmatrix}
    -t_3 z^{-1}\\
    t_1
\end{pmatrix}\}
\end{equation}
We are looking for an element of this kernel, $\alpha_i$, which also solves
\begin{equation}
h_{i}(z)\alpha_{i}(z)=
    \begin{pmatrix}
        0 & t_2\\
        t_2 & 0
    \end{pmatrix}
    \begin{pmatrix}
        -t_3 z^{-1}\\
        t_1
    \end{pmatrix}
    =E(z)
    \begin{pmatrix}
        -t_3 z^{-1}\\
        t_1
    \end{pmatrix}.
\end{equation}
The solution is given by $z=\pm \frac{t_3}{t_1}$ and $E=\mp t_2$. The edge state will be $\alpha_{\pm}=\frac{1}{\sqrt{2}}\begin{pmatrix}
    0&\mp 1 & 1
\end{pmatrix}^T$ since normalization also requires either $|z|\geq 1$ or $|z|\leq 1$ depending on which side the boundary is. This method is equivalent to setting the first sublattice to zero, in which case the wave function must be an eigenvector of $$\begin{pmatrix}
    0 & 0 & 0\\
    0& 0 & t_2 \\
    0& t_2 & 0
\end{pmatrix},$$ and then finding $z$ for which the first sublattice can be zero.

In this model, there is no gap closing when $t_3=t_1$. If these edge states were adiabatically protected, at this value of parameters the edge states would disappear. By looking closely at $h_{2}(z)$, we notice that its eigenstates are
\begin{equation}
    \begin{pmatrix}
        0\\
        -t_3 z^{-1}\\
        t_1
    \end{pmatrix}\text{ and }
    \begin{pmatrix}
        \pm \sqrt{t_1^2+t_3^2}\\
        t_1\\
        t_3 z
    \end{pmatrix},
\end{equation}
none of which are exceptional states (in other words, there is no value of $z$ at which two eigenstates coalesce). Therefore the conditions discussed in \autoref{sec:band crossing} do not hold for this state, and we should not expect there to be a gap closing when edge states appear or disappear.

\section{Zigzag Graphene}\label{app:Haldane}

\begin{figure}[t]
    \centering
    \includegraphics[width=0.75\linewidth]{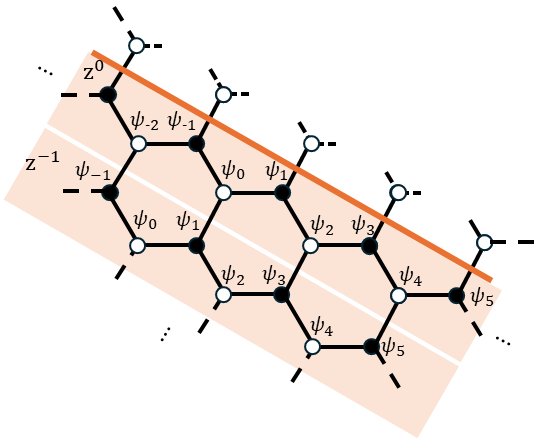}
    \caption{(Color online). A representation of graphene with zigzag boundary condition. The boundary is shown by the orange line. The white circles represent zero amplitude sites, the orange boxes represent the unit cell (the amplitude of the wave function is modulated by $z$ between unit cells).}
    \label{fig:graphene}
\end{figure}

Let us apply the method to a tight binding model of graphene with zigzag boundary condition, shown in \autoref{fig:graphene}. We will also add a staggered potential, so that one sublattice (even index) gets its energy shifted by $\Delta$, and the other (odd index) by $-\Delta$. The hopping parameters are given by $t$.


\begin{figure}[tbp]
     \centering
     \includegraphics[width=1\linewidth]{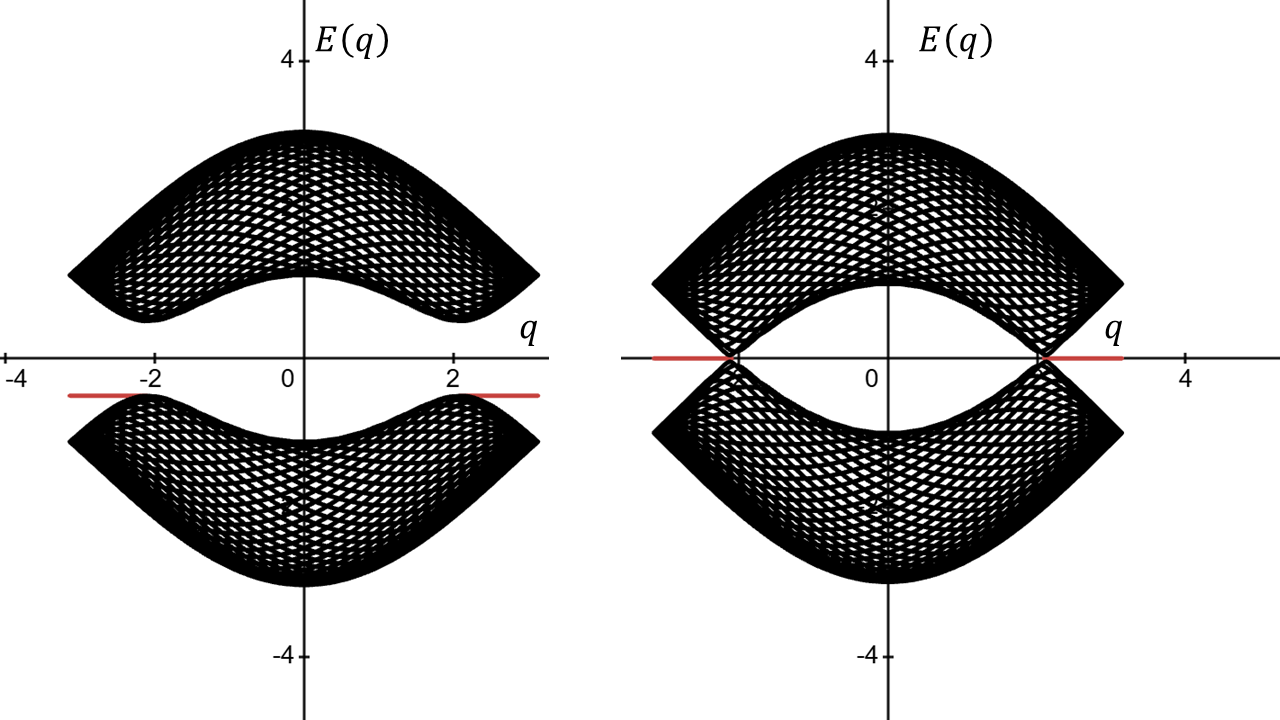}
     \caption{(Color online). The bulk and boundary spectra for the zigzag graphene. Bulk bands are in black, and the edge band is in red. In the left figure, $\frac{\Delta}{t}=\frac{1}{2}$ and in the right $\Delta=0$ .}
     \label{fig:graphene spectrum}
\end{figure}

\begin{figure}[tbp]
    \centering
    \includegraphics[width=.9\linewidth]{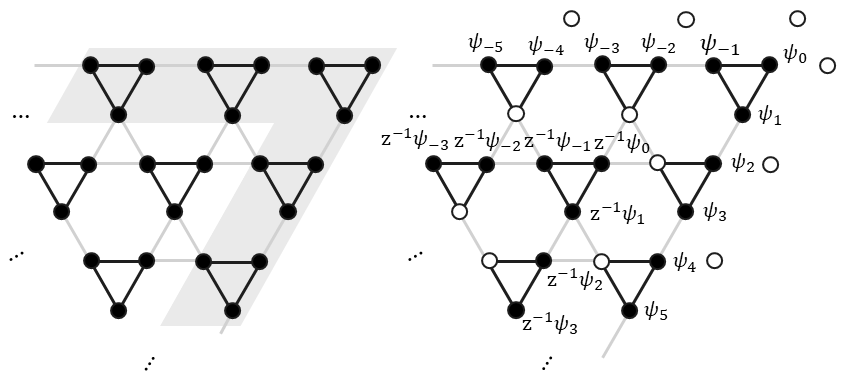}
    \caption{The breathing kagome lattice model. The black lines represent the $t_1$ hoppings, the gray ones represent $t_2$ hoppings. The shaded area represents the unit cell. On the right, we apply the edge chain \textit{ansatz} for this unit cell. }
    \label{fig:Breathing Kagome}
\end{figure}

From \autoref{fig:graphene}, we see that the method imposes every $\psi_{2n}=0$, therefore
\begin{align}
    E\psi_{2n} &= 0  \\
    &= t \psi_{2n-1} + t\psi_{2n+1} + z^{-1}t \psi_{2n+1} + \Delta \psi_{2n} \nonumber\\
    &=t \psi_{2n-1} + t\psi_{2n+1} + z^{-1}t \psi_{2n+1}.
\end{align}
However, the system has a remaining translation symmetry and $\psi_{2n + 1}= e^{iqa}\psi_{2n-1}$ with $q \in [-\pi/a, \pi/a)$, from which it is evident that $\psi_{2n + 1}= e^{iqna}\psi_{1}$ and the above equation reduces to
\begin{equation}
    0 = e^{iq(n-1)a} +  e^{iqna} + z^{-1} e^{iqna},
\end{equation}
from which it follows that
\begin{equation}
    z=\frac{-1}{ e^{-iqa} + 1}.
\end{equation}
The boundary under consideration imposes $|z|>1$ for edge states. $z(q)$ satisfies this condition for $q \in [-\pi/a, -2\pi/(3a)] \cup [2\pi/(3a), \pi/a)$. The energies of the edge states are given by $E(q) = -\Delta$. The bulk and boundary spectrum of the model is presented in \autoref{fig:graphene spectrum}.

\section{Breathing Kagome Lattice}\label{app:Kagome}
\begin{figure}[tbp]
    \centering
    \includegraphics[width=.9\linewidth]{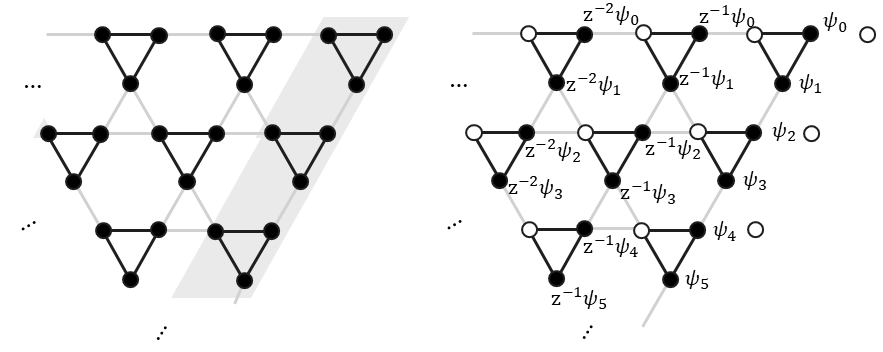}
    \caption{The breathing kagome lattice with another unit cell (shaded area), and the corresponding edge chain \textit{ansatz} (on the right).}
    \label{fig:breathing 2}
\end{figure}

We apply the method to the breathing kagome lattice model illustrated in \autoref{fig:Breathing Kagome}. We start by considering exponential growth towards the corner, in which case we use the unit cell in that same figure.

There is no simple way to write the Bloch Hamiltonian for this unit cell, so it is more convenient to do the method visually, as was done in \autoref{sec:Edge Chain Ansatz}. To start, suppose there are zeros at the appropriate sites, so that the edge chain can be considered independently from the bulk (steps~\ref{item:zero amp} and \ref{item:bloch} in \autoref{sec:Edge Chain Ansatz}). See \autoref{fig:Breathing Kagome} on the right. By applying the Hamiltonian to the zero amplitude sites (step~\ref{item:restrictions} in \autoref{sec:Edge Chain Ansatz}), we get the following conditions on $z$:
\begin{align}
    0&=t_2 z^{-1}\psi_0+t_1 (\psi_{-2}+\psi_{-3}) ,\\
    0&=t_2 z^{-1}\psi_0+t_1 (\psi_{2}+\psi_{3}),\\
    0 &= t_2z^{-1}(\psi_{2i+2}+\psi_{2i+3})+t_1 (\psi_{2i}+\psi_{2i-1}) \text{    for } i\leq -2,\\
    0 &= t_2z^{-1}(\psi_{2i-2}+\psi_{2i-3})+t_1 (\psi_{2i}+\psi_{2i+1}) \text{    for } i\geq 2,
\end{align}
which has the solution
\begin{multline}
    z=-\frac{t_2}{t_1}\frac{\psi_{2i+2}+\psi_{2i+3}}{\psi_{2i}+\psi_{2i-1}}=-\frac{t_2}{t_1}\frac{\psi_{-2i-2}+\psi_{-2i-3}}{\psi_{-2i}+\psi_{-2i+1}}\\
    =-\frac{t_2}{t_1}\frac{\psi_0}{\psi_{\pm2}+\psi_{\pm 3}}.
\end{multline}
By using the fact that $z$ cannot depend on $i$, we get the only valid solution
\begin{equation}
    \psi_i=
    \begin{cases}
        (-\frac{t_1}{t_2})^{|i|/2} \text{ for $i$ even}\\
        0\text{ for $i$ odd}
    \end{cases},
\end{equation}
which is one of the defect-bound solutions of the infinite SSH chain and corresponds to a corner state in the breathing kagome lattice model. The corresponding $z$ is $z = ( \frac{t_2}{t_1} )^2$. The boundary only allows $|z|\geq 1$; therefore a solution exists only if $t_2 > t_1$.

Now let us consider edge states that grow exponentially towards the right side. For these states, we need to consider the unit cell illustrated in the left panel of \autoref{fig:breathing 2}. In the right panel, the sites with zero amplitude are illustrated. Applying step~\ref{item:restrictions}, we obtain the conditions
\begin{equation}
    0=(t_1+z^{-1}t_2)\psi_{2i}.
\end{equation}
If $\psi_{2i}=0$, we obtain, as above, the corner state solution. The other solution, $z=-\frac{t_2}{t_1}$, does not impose anything on the $\psi_i$'s. Therefore, any solution of the edge chain (which here is just an open SSH chain), is also an edge state solution of the breathing kagome lattice, provided $t_1 < t_2$ (otherwise $|z|<1$ which would not be normalizable).

\bibliography{ref}

@article{SSH3,
  title = {Bulk-edge correspondence in the trimer {Su-Schrieffer-Heeger} model},
  author = {Anastasiadis, Adamantios and Styliaris, Georgios and Chaunsali, Rajesh and Theocharis, Georgios and Diakonos, Fotios K.},
  journal = {Phys. Rev. B},
  volume = {106},
  issue = {8},
  pages = {085109},
  numpages = {14},
  year = {2022},
  month = {Aug},
  publisher = {American Physical Society},
  doi = {10.1103/PhysRevB.106.085109},
  url = {https://link.aps.org/doi/10.1103/PhysRevB.106.085109}
}

@mastersthesis{SSH_with_Soliton,
    author = {Maude Bédard},
    title = {Localized electronic states of a centrosymmetric SSH soliton},
    school = {Université de Montréal},
    year = {2022},
}

@unpublished{Nicolas,
    author = {Nicolas Levasseur and Richard MacKenzie and Ilya Iakoub},
    title = {Analysis of the {SSH} chain with a perfectly localized soliton},
    note = {[in progress]}
}

@article{SSH_2D,
  title = {Electronic and topological properties of extended two-dimensional {Su-Schrieffer-Heeger} models and realization of flat edge bands},
  author = {Ma, Hongyu and Zhang, Ze and Fu, Pei-Hao and Wu, Jiansheng and Yu, Xiang-Long},
  journal = {Phys. Rev. B},
  volume = {106},
  issue = {24},
  pages = {245109},
  numpages = {10},
  year = {2022},
  month = {Dec},
  publisher = {American Physical Society},
  doi = {10.1103/PhysRevB.106.245109},
  url = {https://link.aps.org/doi/10.1103/PhysRevB.106.245109}
}

@article{para_Hermitian_1,
author = {Dopico, Froil{\'a}n M. and Noferini, Vanni and Quintana, Mar{\'i}a C. and Van Dooren, Paul},
title = {Para-{Hermitian} Rational Matrices},
journal = {SIAM Journal on Matrix Analysis and Applications},
volume = {45},
number = {4},
pages = {2339-2359},
year = {2024},
doi = {10.1137/24M1678416},

URL = { 
    
        https://doi.org/10.1137/24M1678416
    
    

},
eprint = { 
    
        https://doi.org/10.1137/24M1678416
    
    

}}

@article{Para_Hermitian_2,
  author  = {G.~Barbarino and V.~Noferini},
  title   = {On the {Rellich} eigendecomposition of para-{Hermitian} matrices and the sign characteristics of *-palindromic matrix polynomials},
  journal = {Linear Algebra and its Applications},
  volume  = {664},
  pages   = {1--27},
  year    = {2023},
  doi     = {10.1016/j.laa.2023.05.010}
}

@article{non_Hermitian,
author = {Yuto Ashida and Zongping Gong and Masahito Ueda},
title = {Non-{Hermitian} physics},
journal = {Advances in Physics},
volume = {69},
number = {3},
pages = {249--435},
year = {2020},
publisher = {Taylor \& Francis},
doi = {10.1080/00018732.2021.1876991},


URL = { 
    
        https://doi.org/10.1080/00018732.2021.1876991
    
    

},
eprint = { 
    
        https://doi.org/10.1080/00018732.2021.1876991
    
    

}

}

@article{Higher_order_E.P.,
  title = {Symmetry and Higher-Order Exceptional Points},
  author = {Mandal, Ipsita and Bergholtz, Emil J.},
  journal = {Phys. Rev. Lett.},
  volume = {127},
  issue = {18},
  pages = {186601},
  numpages = {6},
  year = {2021},
  month = {Oct},
  publisher = {American Physical Society},
  doi = {10.1103/PhysRevLett.127.186601},
  url = {https://link.aps.org/doi/10.1103/PhysRevLett.127.186601}
}

@misc{myself,
      title={Bulk-Boundary Correspondence in Semi-Infinite Chains from Sublattice Zeros}, 
      author={Ilya Iakoub and Nicolas Levasseur and Kylian Lionnet and Richard MacKenzie},
      year={2026},
      eprint={2608.01368},
      archivePrefix={arXiv},
      primaryClass={cond-mat.mes-hall},
      url={https://arxiv.org/abs/2608.01368}, 
}

@article{Shemesh,
title = {Common eigenvectors of two matrices},
journal = {Linear Algebra and its Applications},
volume = {62},
pages = {11-18},
year = {1984},
issn = {0024-3795},
doi = {https://doi.org/10.1016/0024-3795(84)90085-5},
url = {https://www.sciencedirect.com/science/article/pii/0024379584900855},
author = {Dan Shemesh}
}

@article{Breathing_Kagome_lattice,
  title = {Exact higher-order bulk-boundary correspondence of corner-localized states},
  author = {Jung, Minwoo and Yu, Yang and Shvets, Gennady},
  journal = {Phys. Rev. B},
  volume = {104},
  issue = {19},
  pages = {195437},
  numpages = {10},
  year = {2021},
  month = {Nov},
  publisher = {American Physical Society},
  doi = {10.1103/PhysRevB.104.195437},
  url = {https://link.aps.org/doi/10.1103/PhysRevB.104.195437}
}

@article{BBH,
author = {Wladimir A. Benalcazar  and B. Andrei Bernevig  and Taylor L. Hughes},
title = {Quantized electric multipole insulators},
journal = {Science},
volume = {357},
number = {6346},
pages = {61-66},
year = {2017},
doi = {10.1126/science.aah6442},
URL = {https://www.science.org/doi/abs/10.1126/science.aah6442},
eprint = {https://www.science.org/doi/pdf/10.1126/science.aah6442}}

@book{Nimrod_Moiseyev,
place={Cambridge},
title={Non-{Hermitian} Quantum Mechanics},
publisher={Cambridge University Press},
author={Moiseyev, Nimrod},
year={2011}}

@book{Kato,
  title={Perturbation Theory for Linear Operators},
  author={Kato, Tosio},
  year={1995},
  edition={2nd},
  publisher={Springer-Verlag},
  address={Berlin, Heidelberg},
  isbn={978-3-540-58661-8},
  doi={10.1007/978-3-540-58661-8}
}

@book{Asboth,
  title={A Short Course on Topological Insulators: Band-structure topology and edge states in one and two dimensions},
  author={Asb{\'o}th, J{\'a}nos K. and Oroszl{\'a}ny, L{\'a}szl{\'o} and P{\'a}lyi, Andr{\'a}s},
  year={2016},
  publisher={Springer},
  series={Lecture Notes in Physics},
  volume={919},
  isbn={978-3-319-25605-4},
  doi={10.1007/978-3-319-25607-8}
}

@article{Haldane,
  title = {Model for a Quantum {Hall} Effect without {Landau} Levels: Condensed-Matter Realization of the ``Parity Anomaly"},
  author = {Haldane, F. D. M.},
  journal = {Phys. Rev. Lett.},
  volume = {61},
  issue = {18},
  pages = {2015--2018},
  numpages = {0},
  year = {1988},
  month = {Oct},
  publisher = {American Physical Society},
  doi = {10.1103/PhysRevLett.61.2015},
  url = {https://link.aps.org/doi/10.1103/PhysRevLett.61.2015}
}

@article{SSH_2D_2,
  title = {Topological edge states in the {Su-Schrieffer-Heeger} model},
  author = {Obana, Daichi and Liu, Feng and Wakabayashi, Katsunori},
  journal = {Phys. Rev. B},
  volume = {100},
  issue = {7},
  pages = {075437},
  numpages = {9},
  year = {2019},
  month = {Aug},
  publisher = {American Physical Society},
  doi = {10.1103/PhysRevB.100.075437},
  url = {https://link.aps.org/doi/10.1103/PhysRevB.100.075437}
}

@article{SSH_2D_3,
title = {Analytic expressions for topologically protected edge states in {Su–Schrieffer–Heeger} model},
journal = {Solid State Communications},
volume = {357},
pages = {114970},
year = {2022},
issn = {0038-1098},
doi = {https://doi.org/10.1016/j.ssc.2022.114970},
url = {https://www.sciencedirect.com/science/article/pii/S0038109822002940},
author = {Jinhong Cheng and Qianru Zhao and Yuqing Zheng and Tie Lin and Xiangjian Meng and Hong Shen and Xudong Wang and Jianlu Wang and Junhao Chu}
}

@book{Wonham,
  author    = {Wonham, W. Murray},
  title     = {Linear Multivariable Control: A Geometric Approach},
  edition   = {3rd},
  series    = {Applications of Mathematics},
  volume    = {10},
  publisher = {Springer-Verlag},
  address   = {New York},
  year      = {1985},
  isbn      = {0-387-96071-6},
  doi       = {10.1007/978-1-4612-1082-5}
}

@article{Kalman,
  author  = {Kalman, R. E.},
  title   = {Mathematical Description of Linear Dynamical Systems},
  journal = {Journal of the Society for Industrial and Applied Mathematics, Series A: Control},
  volume  = {1},
  number  = {2},
  pages   = {152--192},
  year    = {1963},
  doi     = {10.1137/0301010}
}

@article{Hautus,
  author  = {Hautus, M. L. J.},
  title   = {Controllability and Observability Conditions of Linear Autonomous Systems},
  journal = {Indagationes Mathematicae (Proceedings)},
  volume  = {72},
  number  = {5},
  pages   = {443--448},
  year    = {1969}
}

@book{Kailath,
  author    = {Kailath, T.},
  title     = {Linear Systems},
  publisher = {Prentice-Hall},
  address   = {Englewood Cliffs, NJ},
  year      = {1980}
}

@article{Hatsugai,
  author  = {Y.~Hatsugai},
  title   = {Chern number and edge states in the integer quantum Hall effect},
  journal = {Phys. Rev. Lett.},
  volume  = {71},
  pages   = {3697--3700},
  year    = {1993},
  doi     = {10.1103/PhysRevLett.71.3697}
}

@article{Hatsugai_Riemann,
  author  = {Y.~Hatsugai},
  title   = {Edge states in the integer quantum Hall effect and the Riemann surface of the Bloch function},
  journal = {Phys. Rev. B},
  volume  = {48},
  pages   = {11851},
  year    = {1993},
  doi     = {10.1103/PhysRevB.48.11851}
}

@article{Ryu_Hatsugai,
  author  = {S.~Ryu and Y.~Hatsugai},
  title   = {Topological Origin of Zero-Energy Edge States in Particle-Hole Symmetric Systems},
  journal = {Phys. Rev. Lett.},
  volume  = {89},
  pages   = {077002},
  year    = {2002},
  doi     = {10.1103/PhysRevLett.89.077002}
}

@article{Ryu_2010,
   title={Topological insulators and superconductors: tenfold way and dimensional hierarchy},
   volume={12},
   ISSN={1367-2630},
   url={http://dx.doi.org/10.1088/1367-2630/12/6/065010},
   DOI={10.1088/1367-2630/12/6/065010},
   number={6},
   journal={New Journal of Physics},
   publisher={IOP Publishing},
   author={Ryu, Shinsei and Schnyder, Andreas P and Furusaki, Akira and Ludwig, Andreas W W},
   year={2010},
   month=June, pages={065010} }

@article{Kohn1959,
  author  = {W.~Kohn},
  title   = {Analytic Properties of Bloch Waves and Wannier Functions},
  journal = {Phys. Rev.},
  volume  = {115},
  pages   = {809},
  year    = {1959},
  doi     = {10.1103/PhysRev.115.809}
}

@article{BB_Chern,
  author  = {R.~S.~K.~Mong and V.~Shivamoggi},
  title   = {Edge states and the bulk-boundary correspondence in Dirac Hamiltonians},
  journal = {Phys. Rev. B},
  volume  = {83},
  pages   = {125109},
  year    = {2011},
  doi     = {10.1103/PhysRevB.83.125109}
}

@article{Dwivedi_Chua,
  author  = {V.~Dwivedi and V.~Chua},
  title   = {Of bulk and boundaries: Generalized transfer matrices for tight-binding models},
  journal = {Phys. Rev. B},
  volume  = {93},
  pages   = {134304},
  year    = {2016},
  doi     = {10.1103/PhysRevB.93.134304}
}

@article{Alase2017,
  author  = {A.~Alase and E.~Cobanera and G.~Ortiz and L.~Viola},
  title   = {Generalization of {Bloch's} theorem for arbitrary boundary conditions: Theory},
  journal = {Phys. Rev. B},
  volume  = {96},
  pages   = {195133},
  year    = {2017},
  doi     = {10.1103/PhysRevB.96.195133}
}

@article{Balabanov2021,
  author  = {O.~Balabanov and D.~Erkensten and H.~Johannesson},
  title   = {Topology of critical chiral phases: Multiband insulators and superconductors},
  journal = {Phys. Rev. Res.},
  volume  = {3},
  pages   = {043048},
  year    = {2021},
  doi     = {10.1103/PhysRevResearch.3.043048}
}

@article{SSH,
  author  = {W.~P.~Su and J.~R.~Schrieffer and A.~J.~Heeger},
  title   = {Solitons in Polyacetylene},
  journal = {Phys. Rev. Lett.},
  volume  = {42},
  pages   = {1698},
  year    = {1979},
  doi     = {10.1103/PhysRevLett.42.1698}
}

@article{Kunst2017,
  author  = {F.~K.~Kunst and M.~Trescher and E.~J.~Bergholtz},
  title   = {Anatomy of topological surface states: Exact solutions from destructive interference on frustrated lattices},
  journal = {Phys. Rev. B},
  volume  = {96},
  pages   = {085443},
  year    = {2017},
  doi     = {10.1103/PhysRevB.96.085443}
}

@article{Kunst2018,
  author  = {F.~K.~Kunst and G.~van Miert and E.~J.~Bergholtz},
  title   = {Lattice models with exactly solvable topological hinge and corner states},
  journal = {Phys. Rev. B},
  volume  = {97},
  pages   = {241405(R)},
  year    = {2018},
  doi     = {10.1103/PhysRevB.97.241405}
}

@article{Kunst2019,
  author  = {F.~K.~Kunst and G.~van Miert and E.~J.~Bergholtz},
  title   = {Boundaries of boundaries: A systematic approach to lattice models with solvable boundary states of arbitrary codimension},
  journal = {Phys. Rev. B},
  volume  = {99},
  pages   = {085426},
  year    = {2019},
  doi     = {10.1103/PhysRevB.99.085426}
}

@article{LiuWakabayashi2017,
  author  = {F.~Liu and K.~Wakabayashi},
  title   = {Novel Topological Phase with a Zero {Berry} Curvature},
  journal = {Phys. Rev. Lett.},
  volume  = {118},
  pages   = {076803},
  year    = {2017},
  doi     = {10.1103/PhysRevLett.118.076803}
}

@article{Ezawa2018,
  author  = {M.~Ezawa},
  title   = {Higher-Order Topological Insulators and Semimetals on the Breathing Kagome and Pyrochlore Lattices},
  journal = {Phys. Rev. Lett.},
  volume  = {120},
  pages   = {026801},
  year    = {2018},
  doi     = {10.1103/PhysRevLett.120.026801}
}

@article{Benalcazar2017,
  author  = {W.~A.~Benalcazar and B.~A.~Bernevig and T.~L.~Hughes},
  title   = {Electric multipole moments, topological multipole moment pumping, and chiral hinge states in crystalline insulators},
  journal = {Phys. Rev. B},
  volume  = {96},
  pages   = {245115},
  year    = {2017},
  doi     = {10.1103/PhysRevB.96.245115}
}

@article{Schindler2018,
  author  = {F.~Schindler and A.~M.~Cook and M.~G.~Vergniory and Z.~Wang and S.~S.~P.~Parkin and B.~A.~Bernevig and T.~Neupert},
  title   = {Higher-order topological insulators},
  journal = {Sci. Adv.},
  volume  = {4},
  pages   = {eaat0346},
  year    = {2018},
  doi     = {10.1126/sciadv.aat0346}
}

@article{JackiwRebbi,
  author  = {R.~Jackiw and C.~Rebbi},
  title   = {Solitons with fermion number 1/2},
  journal = {Phys. Rev. D},
  volume  = {13},
  pages   = {3398},
  year    = {1976},
  doi     = {10.1103/PhysRevD.13.3398}
}

@article{TeoKane2010,
  author  = {J.~C.~Y.~Teo and C.~L.~Kane},
  title   = {Topological defects and gapless modes in insulators and superconductors},
  journal = {Phys. Rev. B},
  volume  = {82},
  pages   = {115120},
  year    = {2010},
  doi     = {10.1103/PhysRevB.82.115120}
}

@article{Fujita1996,
  author  = {M.~Fujita and K.~Wakabayashi and K.~Nakada and K.~Kusakabe},
  title   = {Peculiar Localized State at Zigzag Graphite Edge},
  journal = {J. Phys. Soc. Jpn.},
  volume  = {65},
  pages   = {1920},
  year    = {1996},
  doi     = {10.1143/JPSJ.65.1920}
}

@article{Semenoff1984,
  author  = {G.~W.~Semenoff},
  title   = {Condensed-Matter Simulation of a Three-Dimensional Anomaly},
  journal = {Phys. Rev. Lett.},
  volume  = {53},
  pages   = {2449},
  year    = {1984},
  doi     = {10.1103/PhysRevLett.53.2449}
}

\end{document}